\documentclass{eas}
\usepackage{graphicx}
\begin{document}

\TitreGlobal{Dark Matter in Spiral Galaxies }

\title{ The Distribution of Dark Matter in Spirals}
\author{Paolo  Salucci}\address{SISSA, Via Beirut 2, Trieste, Italy}

\runningtitle{Dark matter in Spirals }
\setcounter{page}{23}
\index{Salucci, P.}
 

%
\begin{abstract}  

 In the past years    a wealth of  observations 
  allowed to   unravel  the  structural    properties of the Dark Matter Halos  around spirals. First, their rotation  curves  follow   
  an Universal profile (URC) that  can be   described in terms of  an exponential thin stellar disk and a 
      dark halo  with a
constant  density core, whose  relative importance increases  with galaxy luminosity.  
Careful studies of  individual objects, from  dwarfs to giants, reveal that  dark halos have  a core,  whose  size $r_0$   correlates with the central density $\rho_0$. 
  These  properties  are in   serious discrepancy with 
the cuspy density distribution predicted by N-body simulations in collisionless   $\Lambda$CDM  Cosmology. 
\end{abstract}
\maketitle
%
\section{Introduction}

Rotation curves (RC's) of disk galaxies are the best probe for Dark Matter 
(DM) on galactic scale since its discovery. However,   only recently we discovered some crucial
aspects of their mass  distribution by  a large number of high-quality RC's and by 
  improvements in the techniques of the RC mass-modeling. 
\begin{figure}[t]
\begin{center}
\includegraphics[width=.86\textwidth]{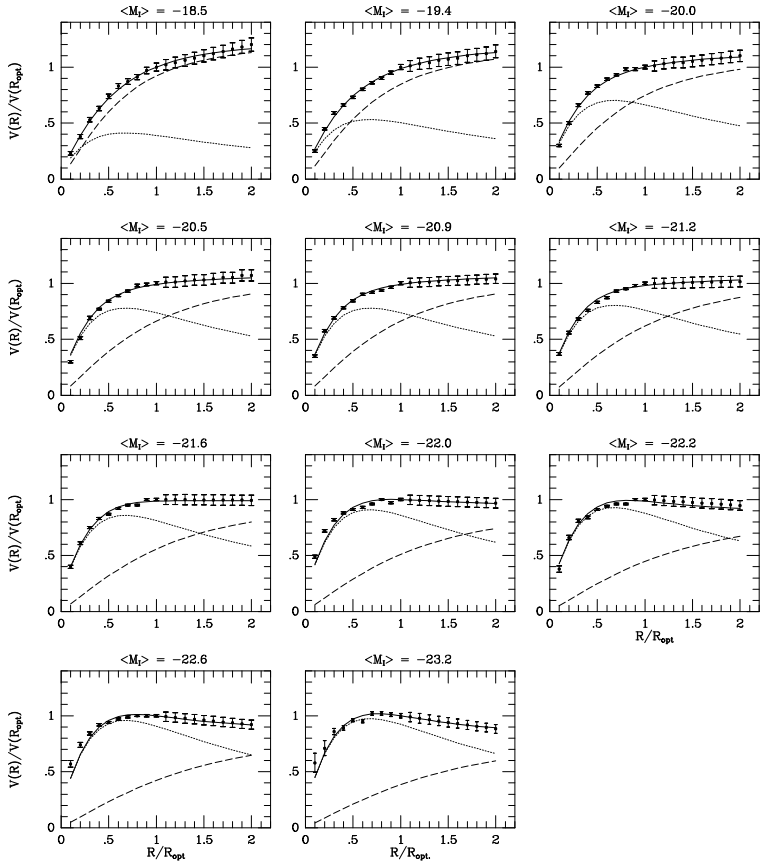}
\vspace {-4.2truecm}
\end{center}
\caption[]{Synthetic rotation curves (circles) and
the
Universal Rotation Curve (solid line).  The  dark/luminous
contributions are indicated with a   dashed/dotted  line.}
\end{figure}
The DM  distribution is usually  assumed according to  one of following different approaches.
     
An empirically one (Persic, Salucci $\&$ Stel, 1996, PSS)  adopts  the simplest  halo {\it velocity}  profile  that (in combination with the stellar disk) reproduces    the Universal Rotation Curve of Spirals  (out to
3-4 $R_D$, $R_{opt} \equiv$ 3.2 disk scale-lenghts $ R_D$): 
\begin{equation}
V_{h,URC}^2(x)= V^2_{opt}\ (1-\beta)\ (1+a^2)\ {x^2 \over (x^2+a^2)}
\end{equation}
where $x \equiv R/R_{opt}$, $a$  is the halo velocity  core radius, in units of
$R_{opt}$   and $\beta \equiv (V_{d}/V_{opt})^2$  is the fractional contribution to 
the circular velocity  of the stellar disk at $R_{opt}$.
This URC-based  profile has  the advantage  of simplicity and,   with suitable choices for 
$\beta$  and  $a$,  it  can  represent  a variety of mass models, including the NFW one. Of course,
it cannot be extrapolated beyond the region traced by the kinematics.     

The  cosmological approach relies on  high--resolution  N--body simulations,  according to which 
 Cold Dark Matter (CDM) halos achieve  the equilibrium density
profile (Navarro, Frenk $\%$ White, 1996, NFW):  
$
\rho_{\rm NFW}(R) = \frac{\rho_s}{(R/r_s)(1+R/r_s)^2}
$
where $r_s$ and $\rho_s$  are the characteristic inner  radius
and density, usually expressed in terms  of virial mass and radius  $r_{\rm vir}$,  $M_{vir}$. 
By   setting   $c \equiv r_{\rm vir}/r_s$,  since   from simulations:      
 $c \simeq  21 (M_{vir}/(10^{11} M_\odot))^{-0.13}$ (Bullock et al. 2001), we have: 
   $V_{\rm NFW}= V_{\rm NFW}(R, M_{vir}, c(M_{vir}))$, and  in detail:
\begin{equation}
V_{\rm NFW}^2(R)= V_{vir}^2 \frac{c}{A(c)} \frac {A(x)}{x}
\end{equation}
where $x \equiv R/r_s$ and $A(x) \equiv \ln (1+x) - x/(1+x)$

The  third approach  adopts   the Burkert profile in order to account for  the  observational evidence at inner radii 
and to converge   to the NFW profile  at outer radii (Salucci $\&$ Burkert, 2000): 
$
\rho_B(R) = \frac{\rho_0 R^3_0}{(R+r_0)(R^2+r_0^2)} 
$ 
with $\rho_0$ and $r_0$ being  the central DM density and the scale radius. Then ($M_0 = 4 \ 1.6 \rho_0 \ r_0^3 $):
\begin{equation}
V^2_B(R) = G {M_0\over{R}} \{ \ln (1 + R/r_0)  -\arctan (R/r_0) + 0.5  \ln  [1
+(R/r_0)^2]\}
\end{equation}

\begin{figure}[t]
\vskip -0.2truecm
\begin{center}
\includegraphics[width=0.52\textwidth]{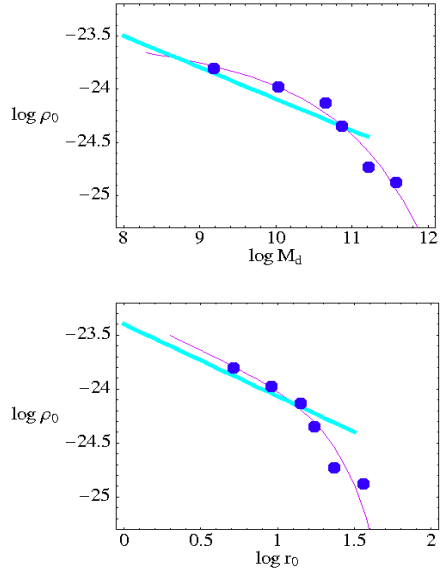}
\vskip -0.6truecm
\end{center}
\caption[]{Central halo density $\rho_0$  (in ${\rm g/cm}^3$)
{\it vs.}  disk mass (in solar units); {\it bottom)} {\it vs.} core radii  (in kpc).  
Details in Salucci $\&$  Burkert, 2001}
\end{figure}
It is important to  stress the following points:  {\it a}) the mass in spirals is
distributed  according to the Inner Baryon
Dominance (IBD) regime (PSS, Salucci and Persic, 1999): there exists  a   "transition radius"
$R_{IBD} \simeq 2  R_d  \ M_d^{1/4}$, with $M_d$ the disk mass in units of $10^{11} M_\odot$. 
For $R \leq R_{IBD}$, (Salucci et al 2000), the luminous matter,  in an exponential thin disk of length-scale $R_D$, 
 accounts  for the whole  gravitating mass.  For  $R > R_{IBD}$, instead,  an additional   dark  component,  distributed unlike the stellar disk,   emerges and it  {\it rapidly}  becomes  dominant.  
{\it b})  the  HI and the bulge contribution to $V(R)$ do not  play a relevant role in the subject 
  of this paper.   

\section{Dark Halos Properties from the Universal Rotation Curve}

PSS have derived,  from  $\sim 20000$ velocity
measurements, relative to  $\sim 900$ rotation curves,   $V_{syn} ({R \over {R_{opt}} }; M_I)$,
 the synthetic  circular velocity of spirals, binned in  intervals of I- magnitudes  (see Fig. 1).
At a fixed luminosity and normalized radius,  the  spiral  RC's  show a  cosmic  variance, with respect to $V_{syn}$, that is  much smaller than  their   radial  variations and than  their luminosity dependence. 
 As result,  spirals sweep a very narrow locus in the RC-~profile/amplitude/luminosity space.  
The whole set of synthetic  RC's    define the Universal Rotation Curve (URC), 
that we analytically represent as  the sum of two terms: {\it a}) the  
exponential thin disk:
\begin{equation}
V^2_{d}(x)=1.28~\beta\ V^2_{opt}~ x^2~(I_0K_0-I_1K_1)|_{1.6x}
\end{equation}
and {\it b)} the  spherical halo  given by eq (1.1). Then: 
  $
 V_{URC}^2(x)=V^2_{h,URC}(x,\beta,a)+ V^2_{d, URC}(x,\beta) 
 $, 
with $a$ and $\beta$ as free parameters,   reproduces the synthetic curves $V_{syn}(R)$ up to their  {\it rms} (i.e.  within
$2\%$) when
$ \beta=\beta (\log V_{opt})$  and  $a =a(\beta)$  as  given in PSS and in  Salucci and Burkert, 2000.  Notice that also (bulge-free) {\it  individual } high-quality  RC are generally well fit by eq. (1.1).
Inside $R_{\rm opt}$ smaller objects have
larger dark-to-stellar  mass ratio:  $M_{\ast}/M_{\rm vir} \simeq 0.2\
(M_{\ast}/2 \times 10^{11} M_{\odot})^{0.75}$.  Scaling  relationships relate the  halo and  disk mass parameters implying  that the densest halos harbor the least massive
disks and  the most inefficient ones to transform the original HI content in present day  stars.  
Notice that the velocity core radius $a$ suggests, but does not prove,  a flat core in the DM  density distribution,
 that can  be unambiguously  revealed only by proper  individual  RC's.  

 \section{The DM  halo density}
   
The mass  structure of spirals is well probed  in objects  with both HI and H$_{\alpha}$ high--quality high-resolution
RC. The existence of the URC and   the  
 Universal properties of cosmological halos  allows us  to concentrate  only in  a  reasonable number of test cases.
The first "absolutely safe" determination of DM halos  density profiles in Spirals was obtained in    Gentile et al 2004,  where  
5 spirals  with  {H{\sc i}} and H$\alpha$ RC's were studied by means of "state of the art"  observational, data analysis and modeling techniques. In each object HI and  H$\alpha$ rotation curves agree very well (where they coexist) and   the combined H$\alpha$ + {H{\sc i}} RC is  smooth, symmetric and 
extended  out to 6-8 disk length-scales.
 The mass distribution  (i.e.$V(R)$) is modeled as the sum of three components: two stellar/gaseous disks
and a spherical dark halo:
$V^2_{model}=V^2_{disk}+V^2_B+ V^2_{ gas}$, with  the halo contribution represented  by the Burkert profile given by eq (1.3).
  Light traces the stellar  mass via a radially constant
mass--to--light ratio. The  gas contribution
$V_{gas}(R)$ is obtained from HI surface brightness and the  distance of the object.
For each galaxy, we determine the values of the structural parameters $\beta $, $r_0$, $\rho_0$ 
by means of a  $\chi ^2$--minimization  of the  velocity model:
$
V^2_{model}(R; \rho_0, \beta , r_0) = V^2_d (R; \beta) + V^2_B (R; \rho_0,  \beta, r_0)+V^2_{gas}(R)
$
   to the (measured) circular velocity $V(R)$, 
subject to  the constraints: $V_{model}(at \ R_{opt}) = V(R_{opt})$  and 
$|dV/dR -dV_{model}/dR|< 0.1 |dV/dR|$ (see Gentile et al. 2004 for details).
This  halo  profile  (+ the exp disk) fits 
the rotation curves extremely well,   with no systematic deviation.  None of the    
 $\sim$ 100 data points  of the five RC's is discrepant   
 at the 3 $\sigma$ level ($\sigma$ is the observational error).
 The stellar I-band mass-to-light ratios lie between
0.5 and 1.8 and  are consistent with population synthesis models.
The presence of cores is clear:  
    $ r_0= (0.7 -- 2.3)
\times~ R_{opt}$ and    $\rho_0 =  (0.4 --
3) \times  10^{-24} g cm^{-3}$. It is worth  noticing the existence of
the relationship among the halo structural parameters  $\rho_0=5\times 10^{-24}r_0^{-2/3}\  exp [-(r_0/27)^2]\  g/cm^3$ 
($r_0$ in kpc, Salucci and Burkert, 2001), crucial   for its implications on  the nature of dark matter       .

\section{High-quality RC's and  NFW halos}

The same objects,  when modeled by means of the   stellar/gas disks + NFW halo components   fail to 
reproduce  the shape of the observed rotation curve. 
Moreover, they show  systematic discrepancy in the predicted velocities: these, in the central parts, 
are too high. In detail:  $10 \%$  of the  measurements  cannot be matched
in any way,  the difference between them and the predictions 
 exceeding  $3 \sigma$  (i.e. three times the   observational error). An other $10\%$ of data  suffer of  a poor
match, the offset is  at the level of  2-3  $\sigma$. This    is a huge model-data  discrepancy: since 1980,  every  mass model  without a central cusp has predicted   circular velocities that resulted    within, at worst,  2  $\sigma$ the observed ones.

\begin{figure}
\begin{center} 
\includegraphics[width=11.6cm,height=6.4cm]{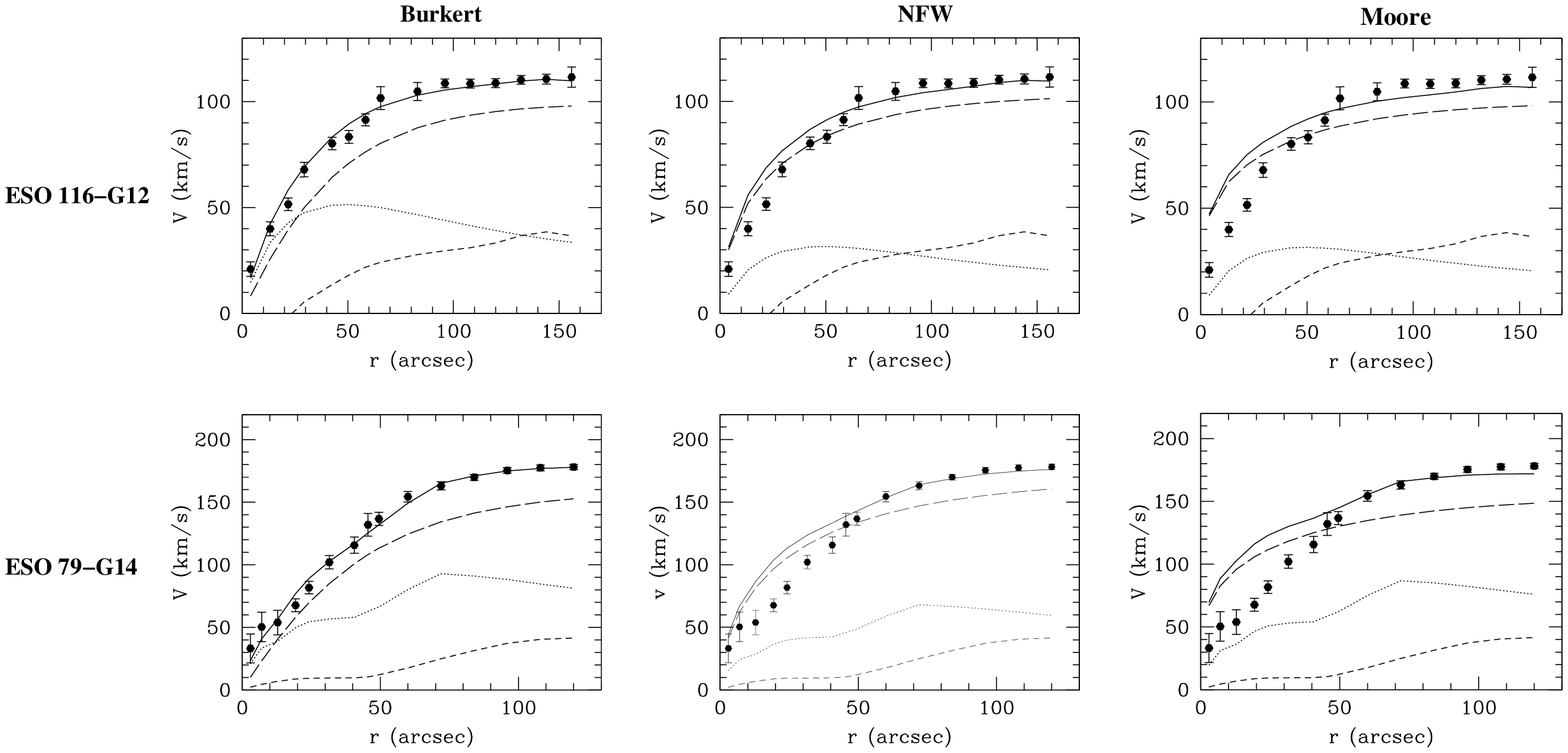}
\end{center}
\vspace{-0.5truecm}
 \caption{   Best-fit models {\it (solid line)}. Also shown the  DM  {\it (long-dashed)}, stellar {\it (dotted)}  and HI {\it (short-dashed)} contributions.  }
\end{figure}
 
Let us notice that,  also if we leave  $c$ as a free parameter,  there is  no appreciable improvement in  the model fits, 
and that, at  galactic scales,  the  actual value of the  density inner slope is  around -1.3, (Navarro et al 2004)
making things even more difficult for standard $\Lambda CDM$. A further shortcoming is that the  resulting NFW disk mass-to-light ratio  turn out to be unacceptably lower than the values  we estimate from  the galaxy color (Gentile et al, 2004).  

Finally, we draw attention  on  the   further  evidence provided by  de Blok and Bosma, 2002, Simon et al 2003, Weldrake et al, 2003, Bolatto et al 2002  about the  serious theory {\it vs}  observations discrepancy in   the DM density distribution, we  show in  Fig 3  a test case.

\begin{figure}[t]
\vspace{-1.2truecm}
 \begin{center}
\includegraphics[width=0.6\textwidth]{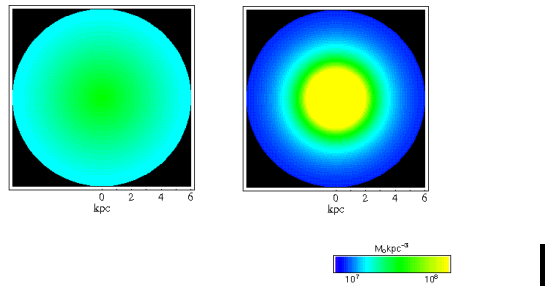}
\end{center}
\vspace{-6.1truecm}
\caption[]{ The density of the  dark halo of   116--G12,and the CDM  predictions {\it  right}  }
\label{eps2}
\end{figure}

\section{Conclusions}

  The mass distribution in spirals underlies  tight relationships
among  their structural   quantities, likely, as the  results of strong feedbacks occurred
during early stages of galaxy formation.   
DM halos around galaxies  have an inner  constant--density region, whose size exceeds the stellar
disk length--scale and  emerge as an
one--parameter family. The order parameter (either the central density or the core
radius)  correlates with the stellar mass. There is no evidence that the density profile  converges,
at large radii, to a $\rho \sim r^{-2}$ (or steeper) profile.  The DM distribution is   
determined by physical parameters, the central core density and the core
radius,  that have no  counterpart in the gravitational 
instability/hierarchical clustering picture.

Solutions for the existence of a region of  ``constant"  density   include a) DM "interacted"  with baryons. Original "cuspy" halos have been  smoothed out b) DM  has a different power spectrum/perturbations evolution than the current Standard Picture c) Dark Matter is a "field", that mimics the effects of a cored halo of particles.
d) the actual dynamical evolution of DM halos,  including  their baryonic content, is more complex than that presently emerging in simulations.


\end{document}